\documentclass[%
 amsmath,amssymb,
 aps,
prd,showpacs,twocolumn,superscriptaddress
]{revtex4-2} 

\usepackage[T1]{fontenc}
\usepackage[utf8]{inputenc}
\usepackage{lmodern}
\usepackage{subfig}
\usepackage{xfrac}
\usepackage{verbatim}
\usepackage{float}
\usepackage{amsthm,amsmath,amssymb}
\usepackage{mathrsfs}
\usepackage{url}
\usepackage{caption}

\usepackage[dvipsnames, usenames]{xcolor}
\definecolor{linkcolor}{rgb}{0.0,0.3,0.5}
\usepackage[hypertexnames=false, unicode, colorlinks=true, linkcolor=linkcolor,
citecolor=linkcolor, filecolor=linkcolor,urlcolor=linkcolor,
pdfusetitle]{hyperref}

\usepackage[all]{hypcap}
\usepackage{graphicx}
\usepackage{xspace}
\usepackage{amssymb}
\usepackage[normalem]{ulem} 
\usepackage{bm} 
\usepackage{enumitem,amssymb}

\usepackage{microtype}
\usepackage[english]{babel}
\usepackage{blindtext}
\usepackage{color}

\def\comment#1{}

\begin{document}

\preprint{APS/123-QED}

\title{Mapping inspiral-merger-ringdown waveforms of binary black holes from black hole perturbation waveforms by machine learning}

\author{Xing-Yu Zhong}
\affiliation{Hangzhou Institute for Advanced Study, University of Chinese Academy of Sciences, Hangzhou 310124, China}
\author{Wen-Biao Han}
\email{corresponding author: wbhan@shao.ac.cn}
\affiliation{Shanghai Astronomical Observatory, Chinese Academy of Sciences, Shanghai 200030, China}
\affiliation{Hangzhou Institute for Advanced Study, University of Chinese Academy of Sciences, Hangzhou 310124, China}
\affiliation{Taiji Laboratory for Gravitational Wave Universe (Beijing/Hangzhou), University of Chinese Academy of Sciences, Beijing 100049, China}
\affiliation{School of Astronomy and Space Science, University of Chinese Academy of Sciences, Beijing 100049, China}
\author{Ling Sun}
\affiliation{OzGrav-ANU, Centre for Gravitational Astrophysics, Research Schools of Physics, and of Astronomy and Astrophysics, The Australian National University, Canberra ACT 2601, Australia}

\date{\today}

\begin{abstract}
Identifying weak gravitational wave signals in noise and estimating the source properties require high-precision waveform templates. Numerical relativity (NR) simulations can provide the most accurate waveforms. However, it is challenging to compute waveform templates in high-dimensional parameter space using NR simulations due to high computational costs. In this work, we implement a novel waveform mapping method, which is an alternative approach to the existing analytical approximations, based on closed-form continuous-time neural networks. This machine-learning-based method greatly improves the efficiency of calculating waveform templates for arbitrary source parameters, such as the binary mass ratio and the spins of component black holes. Based on this method, we present \textit{BHP2NRMLSur}, a class of models (including nonspinning and spin-aligned ones) that maps point-particle black hole perturbation theory waveforms into NR and surrogate waveforms. The nonspinning model provides highly accurate waveforms that match the NR waveforms to the level of $\gtrsim 0.995$. The spin-aligned model reduces the required input parameters and hence improves the efficiency of the waveform generation---it takes a factor of $\sim 50$ less time than existing NR surrogate models to generate $100,000$ waveforms, with a mismatch of $<0.01$ compared to the NR waveforms from the Simulating eXtreme Spacetimes collaboration.
\end{abstract}

\maketitle
\section{Introduction}
The detection of 90 gravitational-wave (GW) signals from compact binary coalescence by the LIGO-Virgo-KAGRA (LVK)~\cite{Aasi_2015, aVirgo, KAGRA} collaboration up to the third observing run (O3) \cite{GWTC-1, GWTC-2, lvkGWTC3} has opened a new era of GW astronomy and significantly expanded our understanding of these compact objects. The increased sensitivity of the detectors in the ongoing fourth observing run (O4) leads to an almost doubled detection rate compared to O3, further increasing the number of observed GW events~\cite{O4rates}. The majority of the observed signals to date are from binary black hole (BBH) mergers.

Generating highly accurate waveform templates is crucial for identifying weak GW signals and estimating their astrophysical properties. Waveforms simulated by numerical relativity (NR) are the most accurate but computationally expensive. Thus, it is impractical to build the full template bank for the whole parameter space of these BBH signals purely relying on NR waveforms. Thousands of NR waveforms with mass ratios $q = m_1/m_2$ ($m_1$ and $m_2$ are the masses of the two-component black holes; $m_1\geqslant m_2$) between 1--10 have been computed and published by the Simulating eXtreme Spacetimes (SXS) collaboration~\cite{SXSweb}. Based on these NR waveforms \cite{nr2013, sxs2019}, several surrogate waveform models are developed to generate waveforms efficiently by reconstructing the underlying phenomenology using a data-driven approach based on NR waveforms~\cite{sxs2019, Varma2019a,Varma2019b, Yoo2022, Yoo2023, Yoo2023spi}. 
These surrogate models are proven to be nearly as accurate as the NR waveforms which they are built upon but allow a much faster generation of waveform templates, and hence become particularly useful in building the template banks for BBH signals.

The next-generation ground-based detectors, such as the Einstein Telescope \cite{ET} and Cosmic Explorer \cite{CE, CE2021}, are expected to reach redshifts beyond $10$ and have significantly improved low-frequency sensitivity, thereby increasing the detection rate of unequal-mass BBH events. In addition, the low-frequency space-based GW detectors Laser Interferometer Space Antenna (LISA)~\cite{lisa}, Taiji~\cite{Taiji}, and Tianqin~\cite{Tianqin} for massive objects are under construction. Intermediate mass ratio inspirals (IMRIs) \cite{2007CQGra..24R.113A} and extreme mass ratio inspirals (EMRIs) \cite{emri2018} are important detection targets for these space-based detectors. Therefore, accurate and efficient modeling of the inspiral-merger-ringdown (IMR) waveforms covering a wide range of mass-ratio systems is essential.

Previous works \cite{rifat2020surrogate,islam2022surrogate} have proposed surrogate models to calibrate point-particle black hole perturbation theory (ppBHPT) waveforms against NR waveforms at different mass ratios by rescaling the amplitude and phase,
\begin{equation}
h_{lm}^{\rm NR}(t,q)\simeq\alpha h^{\rm ppBHPT}_{lm}(t\beta,q),
\end{equation}
where $\alpha$ and $\beta$ are obtained by fitting NR waveforms to polynomials in $1/q$:
\begin{subequations}
    \begin{align}
\alpha(q)&=1+\frac{A^l_\alpha}{q}+\frac{B^l_\alpha}{q^2}+\frac{C^l_\alpha}{q^3}+\frac{D^l_\alpha}{q^4},\label{alpha_l}
\\
\beta(q)&=1+\frac{A^l_\beta}{q}+\frac{B^l_\beta}{q^2}+\frac{C^l_\beta}{q^3}+\frac{D^l_\beta}{q^4},\label{beta_l}
    \end{align}\label{alpha_beta}
\end{subequations}
where $A^l_{\alpha/\beta}$, $B^l_{\alpha/\beta}$, $C^l_{\alpha/\beta}$, and $D^l_{\alpha/\beta}$ are fitting parameters. Recently, the above-mentioned rescaling procedure has been further extended to spinning binaries~\cite{PhysRevD.110.124069}. In this paper, we implement a novel machine-learning-based waveform mapping method to calibrate ppBHPT waveforms against NR waveforms, which is highly efficient and accurate and, more importantly, the flexibility to incorporate additional parameters. Using this method, we map the ppBHPT waveforms to NR waveforms with mass-ratio $q$, aligned individual spins $\chi_1$ and $\chi_2$, and include angular modes of $(2,\,2)$, $(2,\,1)$, $(3,\,3)$, $(3,\,2)$ and $(4,\,4)$. Note that since there are not sufficient NR waveforms available for training purposes, we use SEOBNRv5HM~\cite{ Pompili2023, vandeMeent:2023ols} and $\rm NRHybSur3dq8\_CCE$~\cite{Varma2019a, Yoo2023spi} waveform data for training instead. 

The structure of the paper is organized as follows. In Sec.~\ref{sec:network}, we introduce the machine-learning networks used in this work. In Sec.~
\ref{sec:implementation}, we describe the waveforms, network structure, and the implementation of the training. In Sec.~\ref{sec:models}, we demonstrate the models generated as the training results and the model accuracy by comparing our model-generated waveforms to the surrogate waveforms and NR waveforms from SXS and RIT~\cite{RIT2020}. 
Finally, we outline future directions and conclude in Sec.~
\ref{sec:discussion}.

\section{Methodology}
\label{sec:network}
\begin{figure}[H]
\centering
\begin{tabular}{c}
\includegraphics[width=0.47\textwidth]{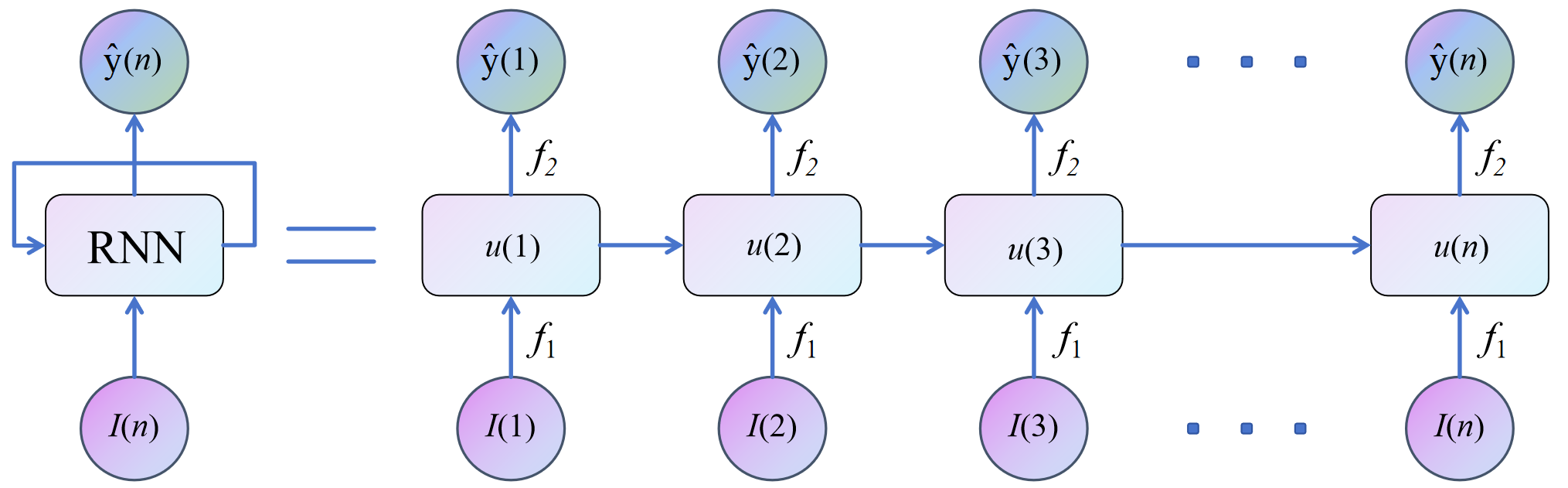}
\end{tabular}
\caption{Schematic diagram of an RNN. The input $I(n)$ and the previous state $u(n-1)$ go into function $f_1$ to generate the current state $u(n)$, and the output $\hat{y}(n)$ is obtained through function $f_2$ acting on $u(n)$. Each step of getting the current state $u(n)$ to output $\hat{y}(n)$ from the input $I(n)$ and the previous hidden state $u(n-1)$ is called a cell.} \label{RNNimag}
\end{figure}

In this work, we rely on the closed-form continuous-time neural network (CfC) \cite{Hasani2021}. Before introducing CfC, it is necessary to review recurrent neural networks (RNNs). The output predictions and computations of RNNs are not only a function of the input at a particular time step but also the past hidden cell state denoted by $u$. Discrete RNNs have the following form:
\begin{eqnarray}
     u(n) &=& f_{1}[u(n-1),I(n), \theta_1], \label{RNN1} \\
     \hat{y}(n) &=& f_{2}[u(n),\theta_2],
    \label{RNN2}
\end{eqnarray}
where $u(n)$ is the hidden cell state at the current step, $f_{1/2}$ is a neural network parametrized by $\theta_{1/2}$ (the same functions and set of parameters are used at every time step in each iteration), $I(n)$ is the input, $u(n-1)$ is the state at last step, and $\hat{y}(n)$ is the output. Figure~\ref{RNNimag} shows the schematic diagram of the RNN in Eqs.~(\ref{RNN1})--(\ref{RNN2}). Note that the input data $[I(1),I(2),...,I(n)]$, hidden states $[u(1),u(2),...,u(n)]$, and output data $[\hat{y}(1),\hat{y}(2),...,\hat{y}(n)]$ are all vectors (or tensors). 

There are two types of RNNs: the discrete-time ones described above and the ones with a continuous-time form. The continuous-time RNNs (CT-RNNs) were proven to approximate any finite time trajectory of a dynamical system \cite{Funahashi1993}. For example, an $m$-dimensional dynamical system can be approximated by a CT-RNN with $m$ units per cell, such that the input at each time step is a $m$-tuple vector $I(t) = [I_1(t), I_2(t),...I_m(t)]$. Assuming that $f_1$ has the following simple form \cite{Funahashi1993}, the CT-RNN will be:
\begin{equation}
    \begin{split}
\dot{u}_i(t) &= -u_i(t)/\tau_i+\sum^m_{j=1}w_{ij}\sigma[u_j(t)]+I_i(t),
    \end{split}\label{CTRNN}
\end{equation}
where the dot symbol denotes differentiation with respect to time $t$, $i=(1,...m)$, $u_i(t)$ is the hidden state of the $i$th unit, $\tau_i$ is the time constant of the $i$th unit, $w_{ij}$ is the connection weights, $I_{i}(t)$ is the input to the $i$th unit, and $\sigma[u_i(t)]$ is the output of $i$th unit. Here, $\sigma$ as the activation function is a nonconstant, bounded, and monotonically increasing $C^1$-sigmoid function: $\sigma(x)=1/(1+e^{-x})$. The vector expression of Eq.~(\ref{CTRNN}) is
\begin{equation}
   \begin{split}
    \dot{\boldsymbol{u}}(t) &= -\boldsymbol{u}(t)/\boldsymbol{\tau}+W{\sigma}[\boldsymbol{u}(t)]+\boldsymbol{I}(t),
    \end{split}\label{CT-RNN}
\end{equation}
where $\boldsymbol{u} = [u_1(t),.... u_m(t)]$, $W = (w_{ij})$ is an $m\times m$ weight matrix. Returning to the general cases, the dimension of the hidden state may not be equal to the dimension of the input, and hence the weight matrix in Eq.~(\ref{CT-RNN}) may not be square. In addition, there are many activation functions. So we rewrite Eq.~(\ref{CT-RNN}) as
\begin{equation}
   \begin{split}
    \dot{\boldsymbol{u}}(t) &= -\boldsymbol{u}(t)/\boldsymbol{\tau}+f_1[\boldsymbol{u}(t), \boldsymbol{I}(t), \theta_1].
    \end{split}\label{CTRNNs}
\end{equation}

\begin{figure}[H]
\centering
\includegraphics[width=0.47\textwidth]{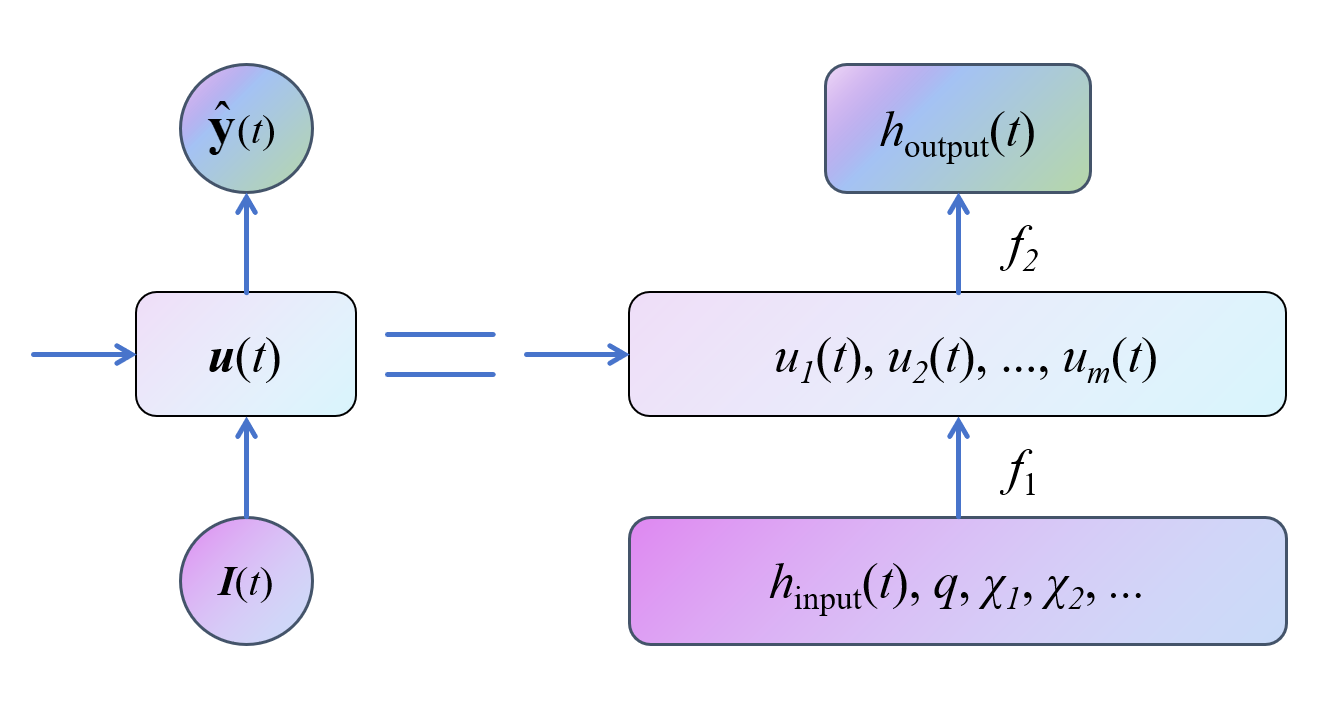}
\caption{Schematic diagram of a cell of the CT-RNN. Such a cell has $m$ units, $[u_1, u_2,...,u_m]$. The input at each time step is a $k$-tuple vector composed of the input waveform and the waveform parameters.
} \label{myRNN}
\end{figure}

In this work, the input is a $k$-tuple vector $I(t)=[h_{\rm input}(t), q, \chi_1, \chi_2, ...]$ at $t$, $h_{\rm input}(t)$ is the ppBHPT waveform and the output $\hat{y}(t) = h_{\rm output}(t) = f_{2}[\boldsymbol{u}(t),\theta_2]$ is the NR approximation waveform at the same time step. At each time, the GW $h\in \mathbb{R}$ and $\boldsymbol{u}=[u_1,u_2,...,u_m]\in \mathcal{D}\subset \mathbb{R}^m$, so we have $f_{1}: \mathbb{R}^m\times\mathbb{R}^{k} \to \mathbb{R}^m$ and $f_2: \mathbb{R}^m\to \mathbb{R}$. Figure~(\ref{myRNN}) shows a schematic diagram of the input, output, and hidden state of the current time step of the neural network cell, from which we can see that $k$ is equal to the number of waveform parameters plus one (the input waveform data). Note that all waveform parameters remain constant over all different time steps of input $I(t)$. Figure~(\ref{myRNN}) shows a single time step in Figure~(\ref{RNNimag}), demonstrating the neural network-based approximation of a single waveform. We illustrate the structural details of the network we use in Sec.~\ref{Modeldetail}.

Reference~\cite{Funahashi1993} provides proof of the theoretical ability of CT-RNNs to approximate any time series. Further, an efficient network training method based on numerical ordinary differential equation (ODE) solvers can be realized using the method described in Reference~\cite{chen2018}. They provide a deep neural network model of neural ODEs. Such a model parametrizes the hidden state's derivative using a neural network. One can then compute the states using a numerical ODE solver and train the network by performing reverse-mode automatic differentiation or by gradient descent through the solver. Then, by concatenating the time-stamp information to the inputs of an RNN, the RNN with neural ODEs can be considered an effective algorithm for modeling the time series data \cite{chen2018}.

Hasani et al. \cite{Hasani2020} provide an alternative formulation to extend the capabilities of CT-RNNs, the liquid time-constant RNNs (LTCs) with varying time-constants coupled to the hidden state: 
\begin{equation}
\boldsymbol{\tau}_{\rm sys}=\frac{\boldsymbol{\tau}}{1+\boldsymbol{\tau} f[\boldsymbol{u}(t), \boldsymbol{I}(t), t, \theta]}.
\end{equation}
The LTC can be implemented by an arbitrarily chosen numerical ODE solver, trained in time by back-propagation and gradient-based gradient optimization algorithm, which is the same as for neural ODEs. In addition, LTCs are proven to have stable and bounded behavior \cite{Lechner2020, Hasani2020}:
\begin{equation}
    \begin{split}
    \dot{\boldsymbol{u}}(t) =& -\{1/\boldsymbol{\tau}+f[\boldsymbol{u}(t), \boldsymbol{I}(t), t, \theta]\}\boldsymbol{u}(t)+\\& f[\boldsymbol{u}(t), \boldsymbol{I}(t), t, \theta]\boldsymbol{A}\\=& -\boldsymbol{u}(t)/\boldsymbol{\tau}_{\rm sys}+ f(\boldsymbol{u}(t), \boldsymbol{I}(t), t, \theta)\boldsymbol{A},
    \end{split}
\end{equation}
where $\boldsymbol{A}$ is a bias vector. The approximate expression of LTCs' closed-form solution is
\begin{equation}
    \begin{split}
    \boldsymbol{u}(t)\approx [\boldsymbol{u}(0)-\boldsymbol{A}]e^{-[1/\boldsymbol{\tau}_{\rm sys}+f(\boldsymbol{u},\boldsymbol{I},\theta)]t}f(-\boldsymbol{u},-\boldsymbol{I},\theta)+\boldsymbol{A}.
    \end{split}\label{c-LTC}
\end{equation}
CfCs are efficiently closed forms of LTCs that can relax the need for complex numerical solvers with well-behaved gradient properties and approximation capabilities. The explicit formation of CfCs is \cite{Hasani2021}:
\begin{equation}
    \begin{aligned}
    \boldsymbol{u}(t)=&\delta[-f(\boldsymbol{u},\boldsymbol{I},\theta_f)t]g(\boldsymbol{u},\boldsymbol{I},\theta_g)+\\
    &\{1-\delta[-f(\boldsymbol{u},-\boldsymbol{I},\theta_f)t]\}a(\boldsymbol{u},-\boldsymbol{I},\theta_a),
    \end{aligned}\label{CfCEq}
\end{equation}
where $\delta$ replaces the exponential term in Eq.~(\ref{c-LTC}). The $\delta$ function is approximately $1$ at $t=0$ and approaches $0$ when $t\to\infty$, much smoother than the exponential decay. The $1-\delta(\cdot)$ term plays the gating role in avoiding problems with a disappearing gradient. Here, we replace $A$ with another neural network$a$ and add an independent network $g$ to improve the model's flexibility. Since CfCs are the better-performed variant of RNNs, we use them to find the mapping relation between ppBHPT and NR waveforms.

\begin{figure*}
\centering
\includegraphics[width=0.9\textwidth]{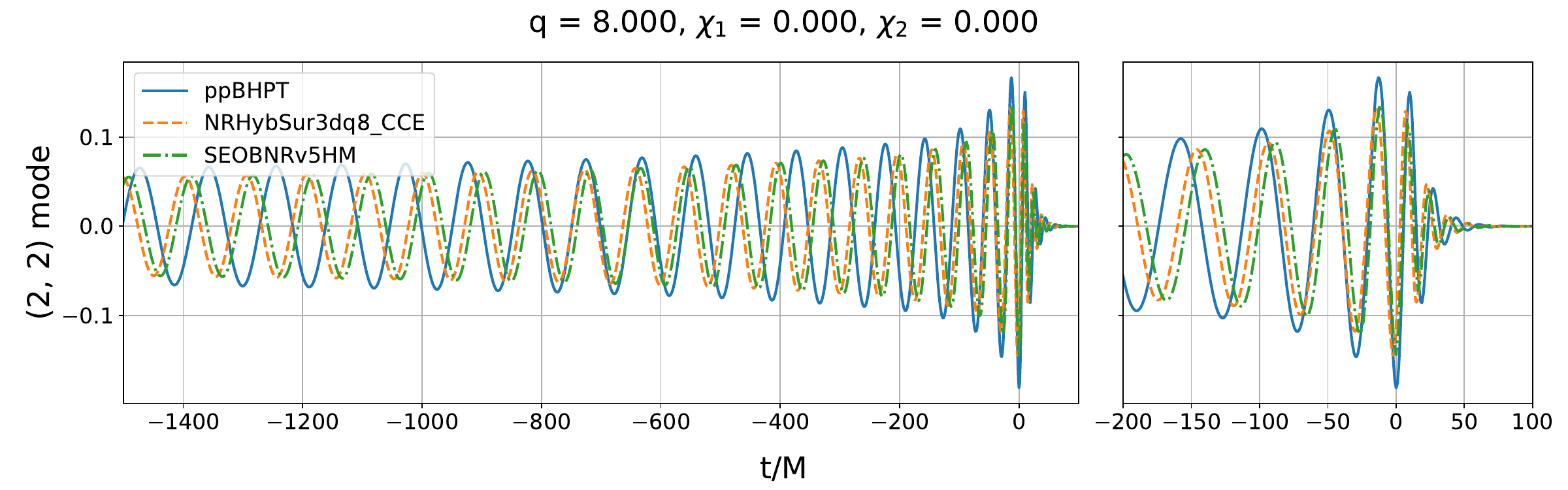}
\captionsetup{justification=raggedright, singlelinecheck=false}
\caption{The $(2,2)$ modes of the ppBHPT (solid blue), $\rm NRHybSur3dq8\_CCE$ (dashed orange) and SEOBNRv5HM (dashed green) waveforms for a system with mass ratio of $q=8$ and spins of two BHs, $\chi_1=\chi_2=0$. The right panel is a zoomed-in view around the merger time.} \label{SXS-bhp}
\end{figure*}

\section{Implementation}
\label{sec:implementation}
\subsection{Waveform data}
In the ppBHPT approximation, the small object in the binary is modeled as a point particle without internal structure and with a mass of $m_2$, moving in the background spacetime of the massive black hole with a mass of $m_1$. As the small object perturbs the background spacetime, GWs are excited. 

The ppBHPT waveforms can be computed by solving the Teukolsky equation \cite{Teukolsky1973, Sundararajan2007, Zengino2011} with the particle trajectory determined source term.  The solution of the equation is directly related to the $4$th Weyl curvature component, $\psi_4$, which can be written as
\begin{equation}
\begin{split}
\psi_4=\frac{1}{2}\left(\frac{\partial^2h_+}{\partial t^2}-i\frac{\partial^2h_{\times}}{\partial t^2}\right),
\end{split}
\end{equation}
where $h_{+}$ and $h_{\times}$ are the two GW polarization states.

We compute the training ppBHPT waveforms based on interpolation data provided by Reference~\cite{islam2022surrogate}. The dataset is generated by interpolating $41$ nonspinning ppBHPT waveforms with the logarithmic scale of $q\in[2.5,\,10000]$.
Since a large number of target NR waveforms required in the training are still lacking, we generate the target data using NR surrogate models, $\rm NRHybSur3dq8\_CCE$ \cite{Varma2019a, Yoo2023spi} and SEOBNRv5HM~\cite{ Pompili2023, vandeMeent:2023ols}. For this reason, two target data-based models ($\rm NRHybSur3dq8\_CCE$ based and SEOBNRv5HM based) are trained based on whether $\rm NRHybSur3dq8\_CCE$ or SEOBNRv5HM generates the target waveform data. All waveforms are aligned, and their peaks occur simultaneously at $t=0$. In addition, we set the phase $\phi= 0$ at the start of the waveforms.

The surrogate model $\rm NRHybSur3dq8\_CCE$ is a hybrid waveform model, trained on $102$ waveforms built from Cauchy-characteristic evolution (CCE) waveforms and post-Newtonian (PN) and effective-one-body (EOB) waveforms \cite{Pompili2023,vandeMeent:2023ols}. SEOBNRv5HM~\cite{ Pompili2023, vandeMeent:2023ols} is an IMR gravitational waveform model for quasicircular, spinning, nonprecessing binary black holes within the EOB formalism. It is calibrated to larger mass ratios and spins using $442$ NR simulations and $13$ BHP waveforms.

There are significant phase and amplitude differences between ppBHPT waveform and NR or NR surrogate waveforms, especially when the mass-ratio $q$ is small. To demonstrate the ability of the calibration capability of our model \textit{BHP2NRMLSur}, Figure~\ref{SXS-bhp} shows the $(2,2)$ modes of the ppBHPT, $\rm NRHybSur3dq8\_CCE$ and SEOBNRv5HM waveforms for comparison with later \textit{BHP2NRMLSur} results.

\subsection{Network structure}\label{Modeldetail}
\begin{figure}[htb!]
\centering
\includegraphics[width=0.47\textwidth]{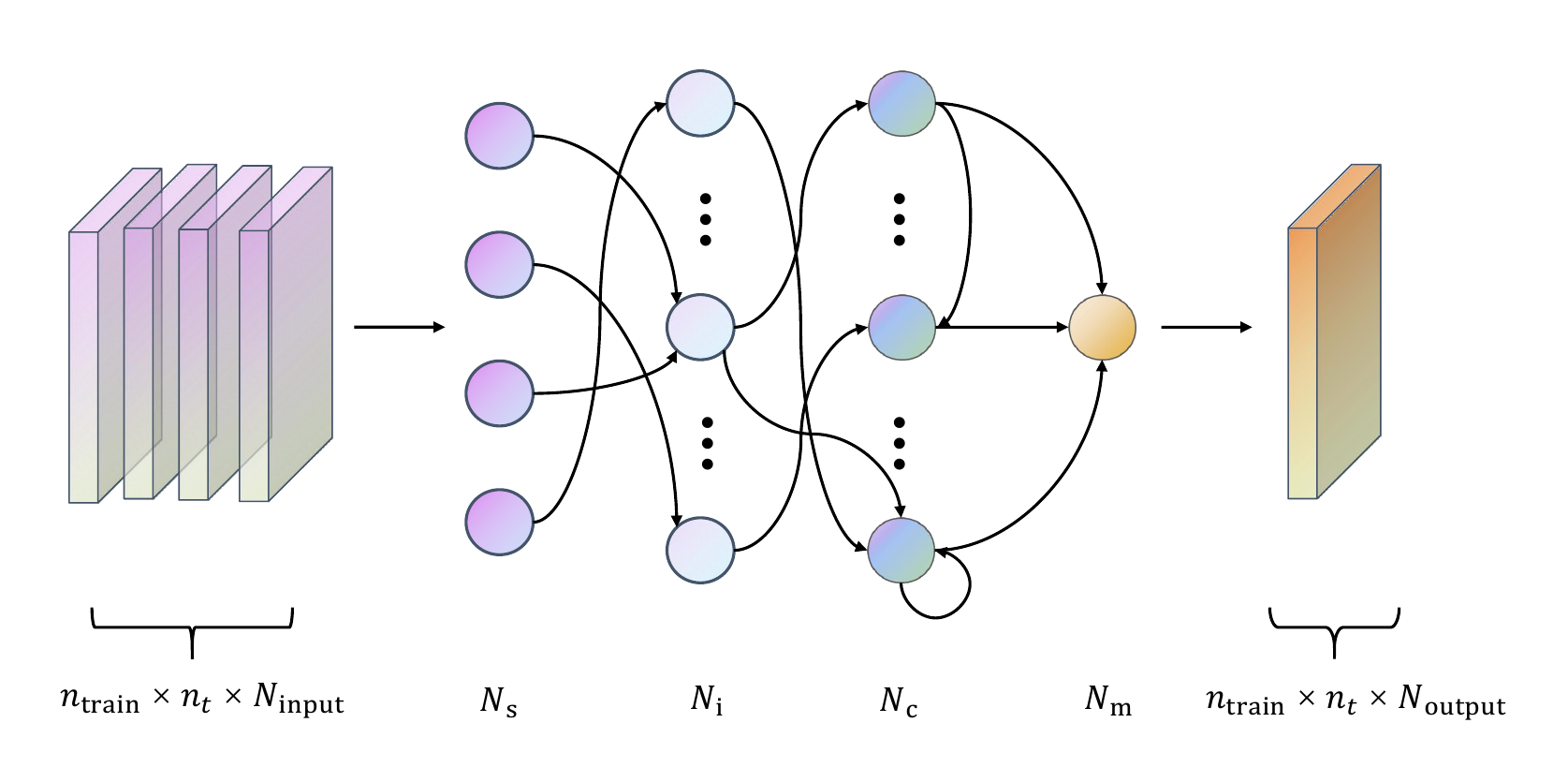}
\captionsetup{justification=raggedright, singlelinecheck=false}
\caption{The NCPs have $N_{\rm s}$ sensory neurons, $N_{\rm i}$ inter neurons, $N_{\rm c}$ command neurons, and $N_{\rm m}$ motor neurons. On the input and output ends, $n_{\rm train}$ is the number of the input training waveforms, $N_{\rm input}$ and $N_{\rm output}$ are the dimensions of the input $I(t)$ and the output $\hat{y}(t)$, respectively, and $n_t$ is the number of time steps. In the process, we have $N_{\rm s}=N_{\rm input}$, $N_{\rm m}=N_{\rm output}$, and $N_{\rm c}=40\%(N_{\rm i}+N_{\rm c})$. } \label{NCPs}
\end{figure}

\begin{figure}[htb!]
\centering
\includegraphics[width=0.47\textwidth]{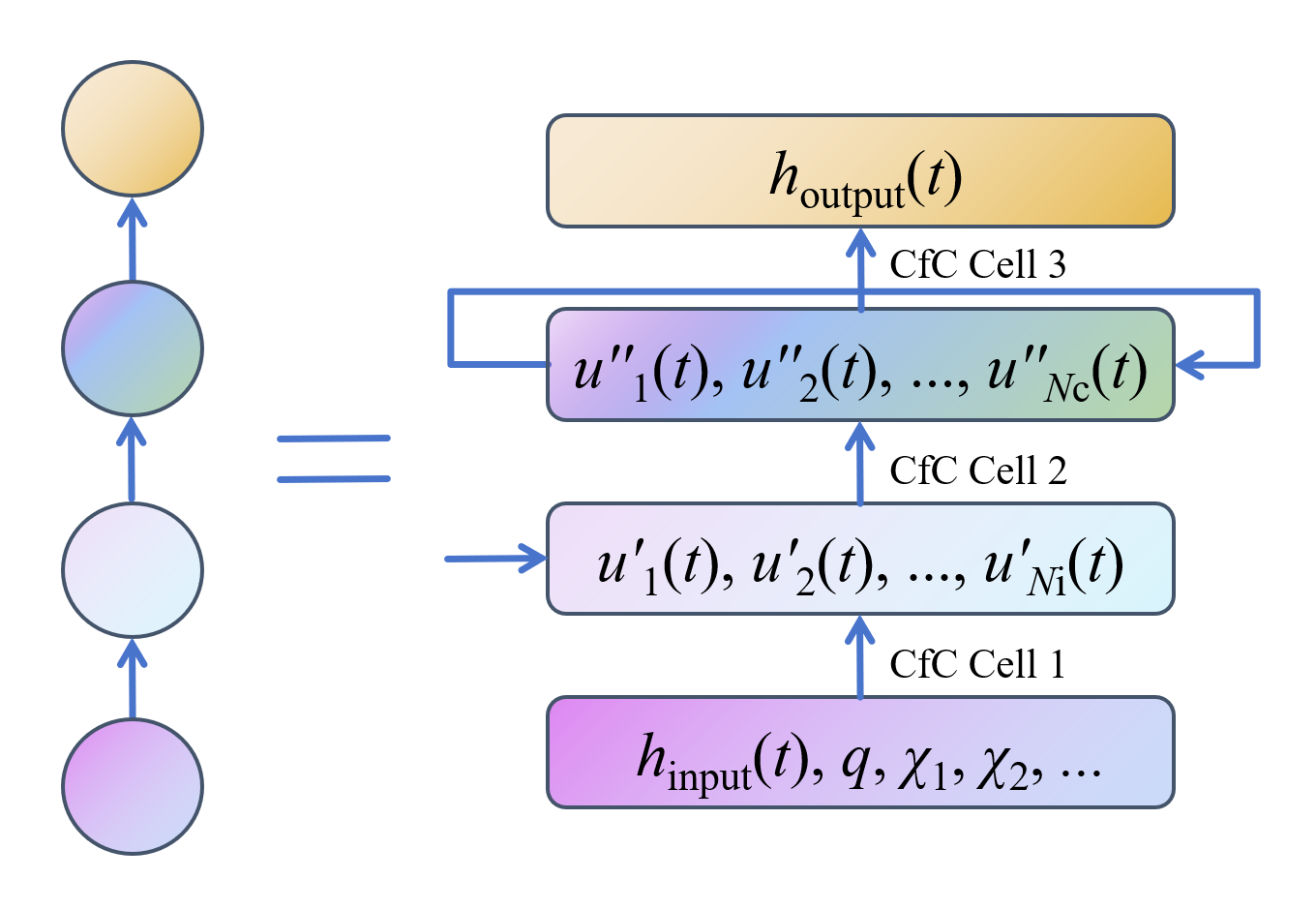}
\captionsetup{justification=raggedright, singlelinecheck=false}
\caption{The process from input through neurons to output at each time step. Circles of different colors correspond to neurons with the same color in Fig~\ref{NCPs}. The ``$'$'' and ``$''$'' mean hidden state of different layers and the ``CfC Cell'' indicate the layer-to-layer transfer relationship in Eq.~(\ref{CfCEq}).} \label{dNCPs}
\end{figure}

To improve the model's expressiveness in mapping waveforms, we use a CfC network with a multineuron neural circuit strategy (NCP) ~\cite{Lechner2020}. The total $N_{\rm tot}$ neurons ($N_{\rm tot}=16$ or $64$ in our model) of NCP are divided into four layers with $N_{\rm s}$ sensory neurons, $N_{\rm i}$ interneurons, $N_{\rm c}$ command neurons, and $N_{\rm m}$ motor neurons, respectively. NCP has a nonlinear and sparse transmission mechanism shown in Figs.~\ref{NCPs} and \ref{dNCPs}. The $50\%$ dropout of feedforward connections between neurons and highly recurrent connections among command neurons achieve sparsity and nonlinearity and reduce the trainable parameter space. The activation function we use is LeCun's Tanh, $f(x)=1.7519\tanh(\frac{2}{3}x)$. It adjusts the slope of the $\tanh$ function to produce a larger gradient near the origin and helps the neural network to speed up the learning process. 


Before training, we need to preprocess the waveform data. The GWs propagating in direction $(\theta,\, \phi)$ can be decomposed into spin-weighted spherical harmonic modes
\begin{equation}
\begin{split}
h_+-ih_{\times}(t,\theta,\phi)=\sum^{\infty}_{l=2}\sum^{l}_{m=-l}h_{lm}(t)_{-2}Y_{lm}(\theta,\phi),
\end{split}
\end{equation}
where $h_{lm}(t)$ is the strain of the $lm$ harmonic mode, and $_{-2}Y_{lm}(\theta,\phi)$ is the spin-weighted spherical harmonics. For the training, it is useful to decompose the waveforms into amplitude and phase components:
\begin{equation}
{h}_{lm}(t) = {A}(t)e^{-i\phi(t)},
\end{equation}
where the amplitude can be written as $A=|h_{lm}|$, and phase is $\phi = \arctan[-{\rm Im}(h_{lm})/{\rm Re}(h_{lm})]$. Our models transform the amplitude ($A_{\rm p}$) and phase ($\phi_{\rm p}$) of the ppBHPT into the prediction amplitude and phase, $A_{\rm P}$ and $\phi_{\rm P}$, respectively. We can also consider the model as a function acting on the input data in the following form:
\begin{equation}
\alpha_{\rm P} = {\rm CfC}^{lm}(\alpha_{\rm p})
\end{equation}
where $\alpha=A,~\phi$.

For model optimization during training, we use the mean squared error as the loss function, which measures the mean squared $\ell^2$-norm between each element in the prediction $(A/\phi)_{\rm P}$ and target $(A/\phi)_{\rm NR}$. We chose the Adam optimiser~\cite{adam2014} to optimize the network trainable parameters at each step of the training process.

\section{Machine-learning based models}
\label{sec:models}
We use \textit{BHP2NRMLSur} to generate IMR waveforms, mapping from the ppBHPT waveforms. In this section, we describe the details of the nonspinning and aligned-spinning models of \textit{BHP2NRMLSur}.
We also present the model accuracy and the efficiency of waveform generation. The quantity of training waveforms and model parameters is crucial for model construction. Table~\ref{Table_model} presents the total number of training waveforms, neurons, and parameters for each type of model. The number of model parameters depends on the number of neurons and waveform parameters. For the nonspinning model, there are a total of $4.2\times10^3$ parameters, including $3.4\times10^3$ trainable parameters and $784$ non-trainable parameters. For the aligned-spinning models, there are $1.62\times10^4$ parameters in total, consisting of $1.3\times10^4$ trainable parameters and $3.2\times10^3$ non-trainable parameters. 

\begin{table}[H]
\caption{\label{Table_model}Total number of training waveforms, model neurons, and parameters in \textit{BHP2NRMLSur}. Aligned-spinning-A and B represent NRHybSur3dq8$\_$CCE-based and the SEOBNRv5HM-based model, respectively.}
\centering
\begin{tabular}{c|c|c|c}
\hline\hline
 &Training waveforms&Neurons&Params\\
 \hline
 nonspinning&$1000$&$16\times4$&$3.4\times10^3$
 \\
 \hline
 Aligned-spinning-A&$30\times30\times30$&$64$&$1.62\times10^4$
 \\
 \hline
 Aligned-spinning-B&$50\times50\times50$&$64$&$1.62\times10^4$
 \\
\hline\hline
\end{tabular}
\end{table}

\subsection{Nonspinning model}
The nonspinning model has the following form:
\begin{equation}
{\rm CfC}^{lm}_\alpha[\alpha(q)]=\left(1+\sum_{n=1}^{n_{\rm max}}\frac{{\rm CfC}^{lmn}_\alpha}{q^{n}}\right)\alpha(q),\label{chi0}
\end{equation}
where $n_{\rm max} = 4$ is chosen here, which means that the nonspinning model consists of four CfC networks, and each CfC network has $16$ neurons ($N_{\rm tot}=16$). It is worth noting that when $q\to\infty$, the mapping function of Eq.~(\ref{chi0}) will approximate the identity mapping and guarantee the accuracy of the mapping waveform in extreme mass ratio case.

For the training, we generate $1000$ ppBHPT waveforms with uniformly sampled mass-ratio $q\in[3,8]$ as the input data $\alpha(q)$ (i.e., the amplitude and phase of waveforms), and the corresponding target data are generated by $\rm NRHybSur3dq8\_CCE$. Here, the waveforms include $(2, 2)$, $(2, 1)$, $(3, 3)$ $(3, 2)$ and $(4, 4)$ harmonic modes. The duration 
of each waveform is $t \in [-2000M, 110M]$ and time step is ${\rm d} t=1M$. 

To evaluate the accuracy of the \textit{BHP2NRMLSur} waveforms, we quantify the discrepancy (mismatch) between a large set of waveforms generated by \textit{BHP2NRMLSur} and NR surrogate ($\rm NRHybSur3dq8\_CCE$, SEOBNRv5HM) or NR (SXS, RIT) waveforms. The match is defined as
\begin{equation}
\label{meq}
\mathcal{O}={\rm max}\left[\frac{\left<h_1|h_2\right>}{\sqrt{\left<h_1|h_1\right>\left<h_2|h_2\right>}}\right],
\end{equation}
where the ``max'' stands for the maximum overlap given by optimizing over time 
and phase shifts, and the $\left<h_1|h_2\right>$ defined as
\begin{align}
\label{snreq}
\left<{h}_1|{h}_2\right>=2{\rm Re}\int^{\infty}_0\frac{\tilde{h}_1^*(f)\tilde{h}_2+\tilde{h}_1(f)\tilde{h}_2^*(f)}{S_n(f)}{\rm d}f,
\end{align}
where the tilde symbol denotes the Fourier transform of the strain $h$, the $``*"$ denotes the complex conjugate, and $S_n(f)$ is the one-sided noise power spectral density (PSD) of the GW detector. $\mathcal{O}_{\rm IMR}$ and $\mathcal{O}_{\rm 
 I}$ denote the matches of the IMR waveforms and the waveforms before merging, respectively.

\begin{figure}[htb!]
    \centering
    \includegraphics[width=0.49\textwidth]{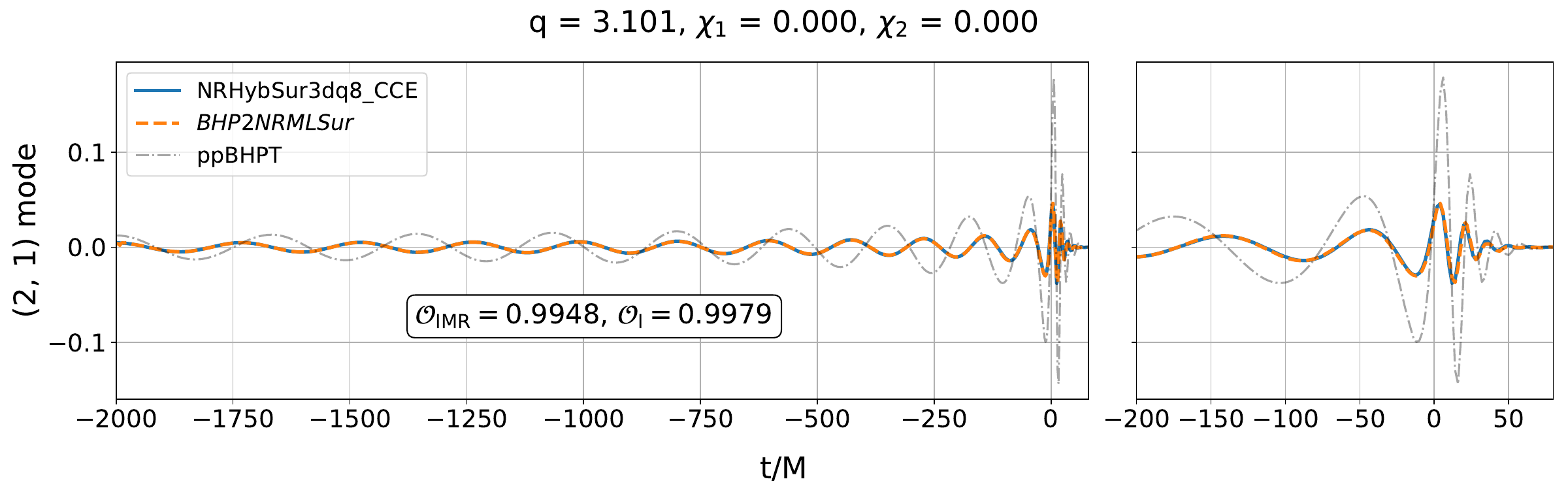}
    \includegraphics[width=0.49\textwidth]{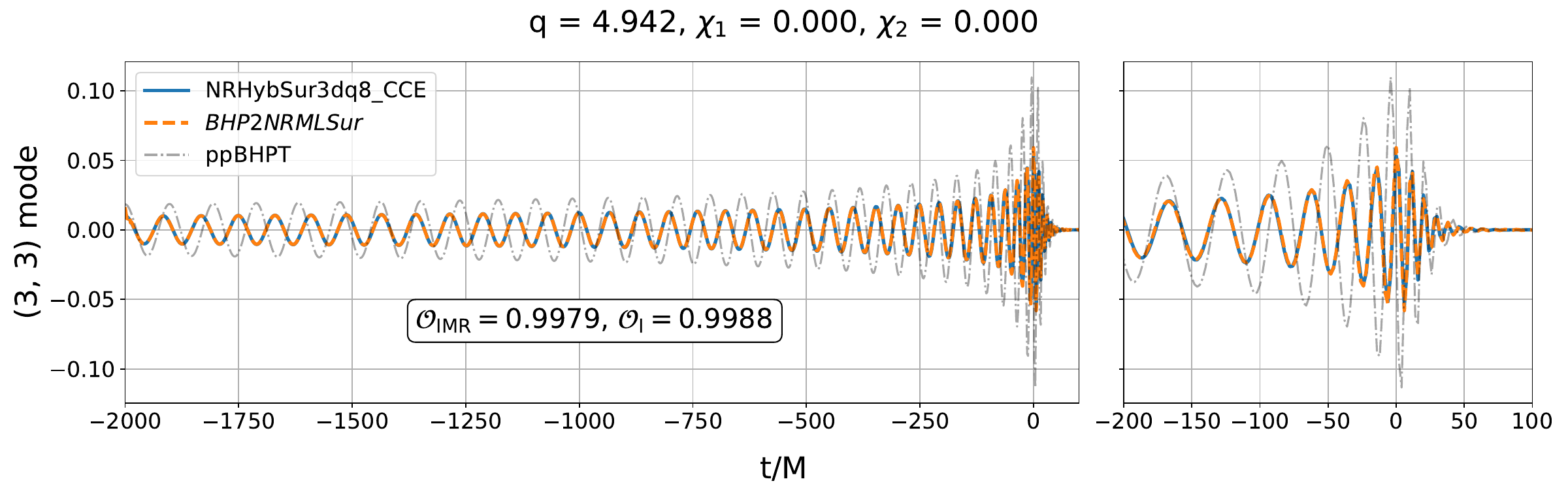}
    \includegraphics[width=0.49\textwidth]{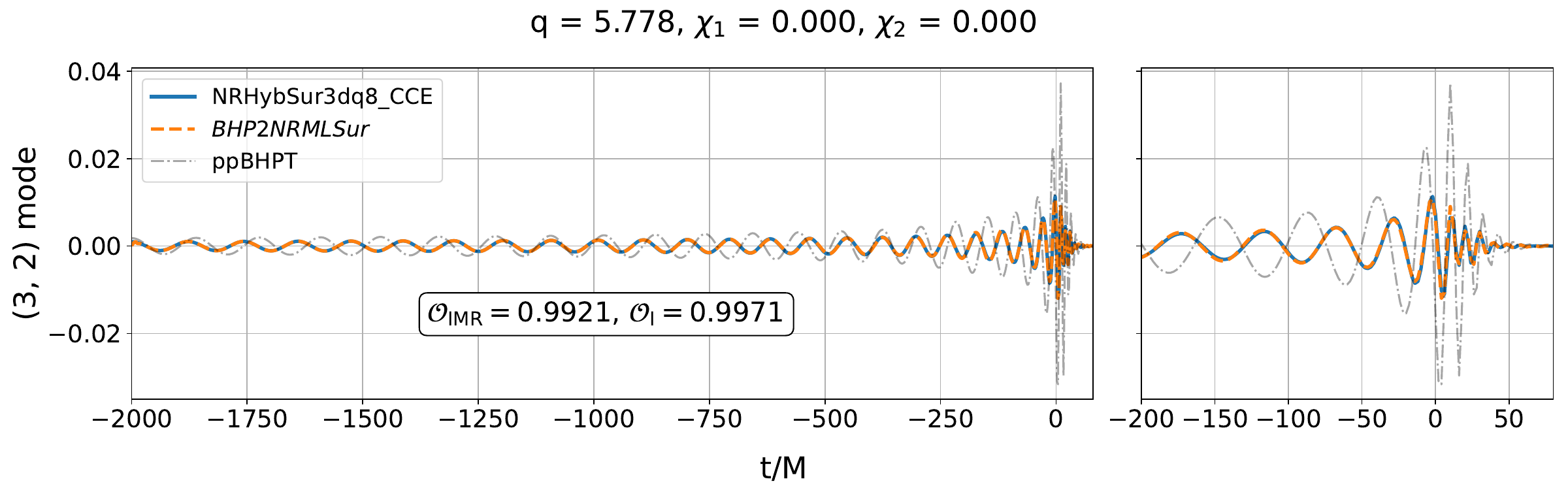}
    \includegraphics[width=0.49\textwidth]{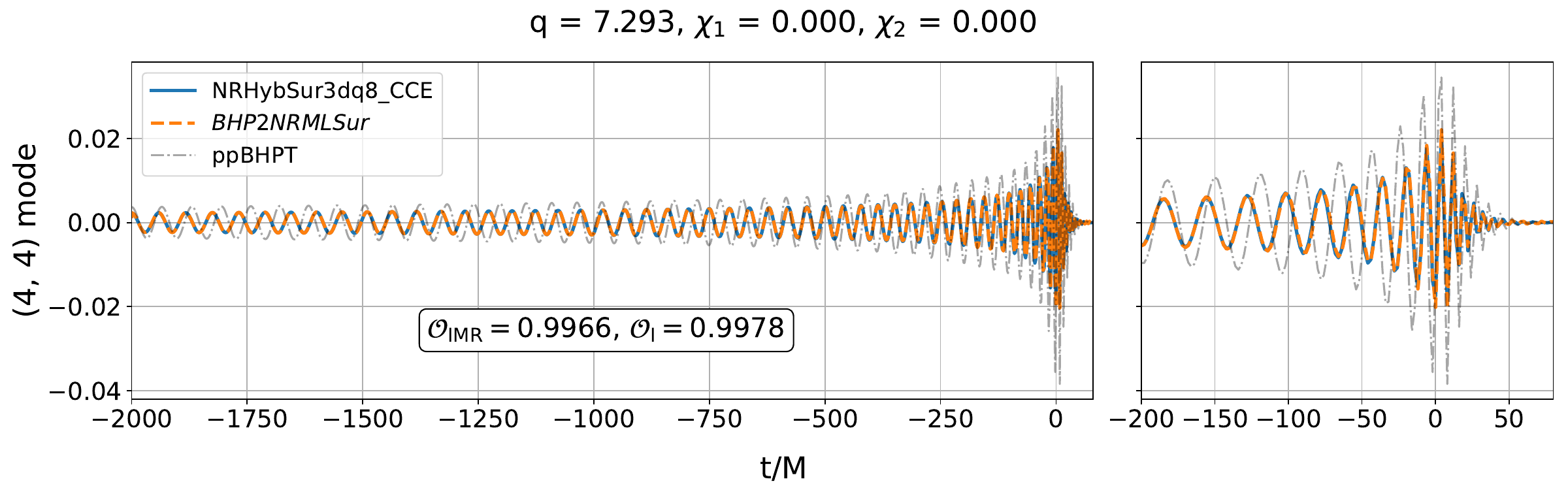}
    \captionsetup{justification=raggedright, singlelinecheck=false}
    \caption{Harmonic modes (orange dashed curves) generated by the non - spinning \textit{BHP2NRMLSur} model and the corresponding $\rm NRHybSur3dq8\_CCE$ waveforms (blue solid curves) and the uncalibrated ppBHPT waveforms are represented by light black dashed lines. From top to bottom are $(2,1)$, $(3,3)$, $(3,2)$ and $(4,4)$ modes.}
    \label{cce_mode}
\end{figure}

\begin{figure}[htb!]
\centering
\includegraphics[width=0.47\textwidth]{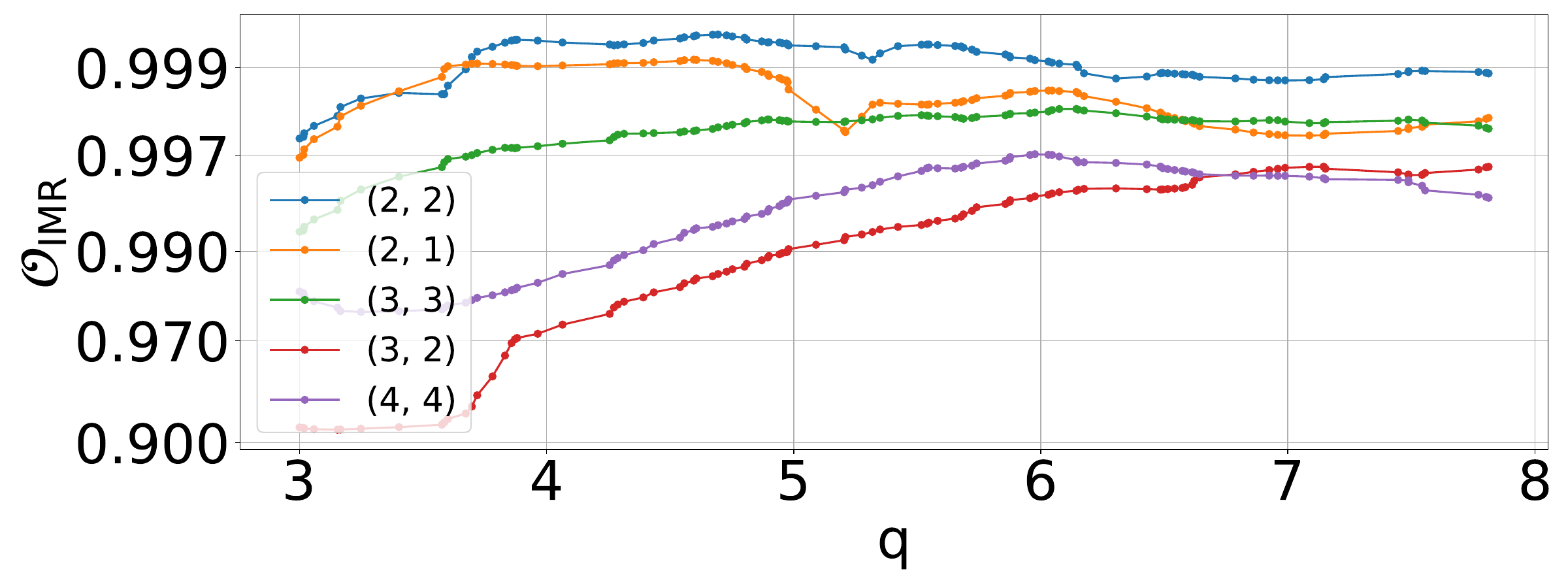}
\captionsetup{justification=raggedright, singlelinecheck=false}
\caption{The matches $\mathcal{O}_{\rm IMR}$ of the harmonic modes of \textit{BHP2NRMLSur} and the corresponding modes of $\rm NRHybSur3dq8\_CCE$.} \label{cce-bhp}
\end{figure}

Figure~\ref{cce_mode} shows the \textit{BHP2NRMLSur} generated harmonic modes and the corresponding modes of $\rm NRHybSur3dq8\_CCE$. Then, we use Eq.~(\ref{meq}) to calculate the matches, and the results are shown in Figure~\ref{cce-bhp}, where each mode is evaluated with 100 waveforms with the mass-ratio $q$ randomly sampled from a uniform distribution in the interval $q\in [3,8]$. For the harmonic modes $(2, 2)$, $(2, 1)$ and $(3, 3)$, the matches between our model-generated waveforms and the waveforms of $\rm NRHybSur3dq8\_CCE$ are pretty high, at the level of $\gtrsim 0.99$, which demonstrates that \textit{BHP2NRMLSur} can obtain highly accurate mapping waveforms from ppBHPT waveforms.

%
%
%
%
%
%

\begin{figure}[htb!]
\centering
\includegraphics[width=0.49\textwidth]{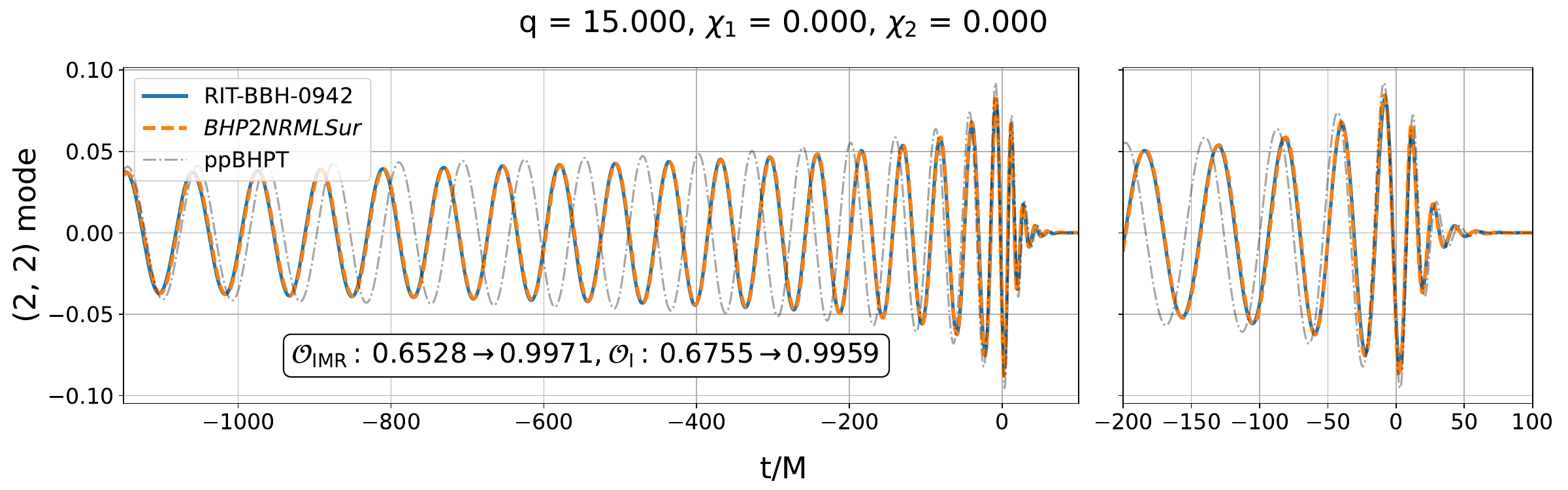}
\includegraphics[width=0.49\textwidth]{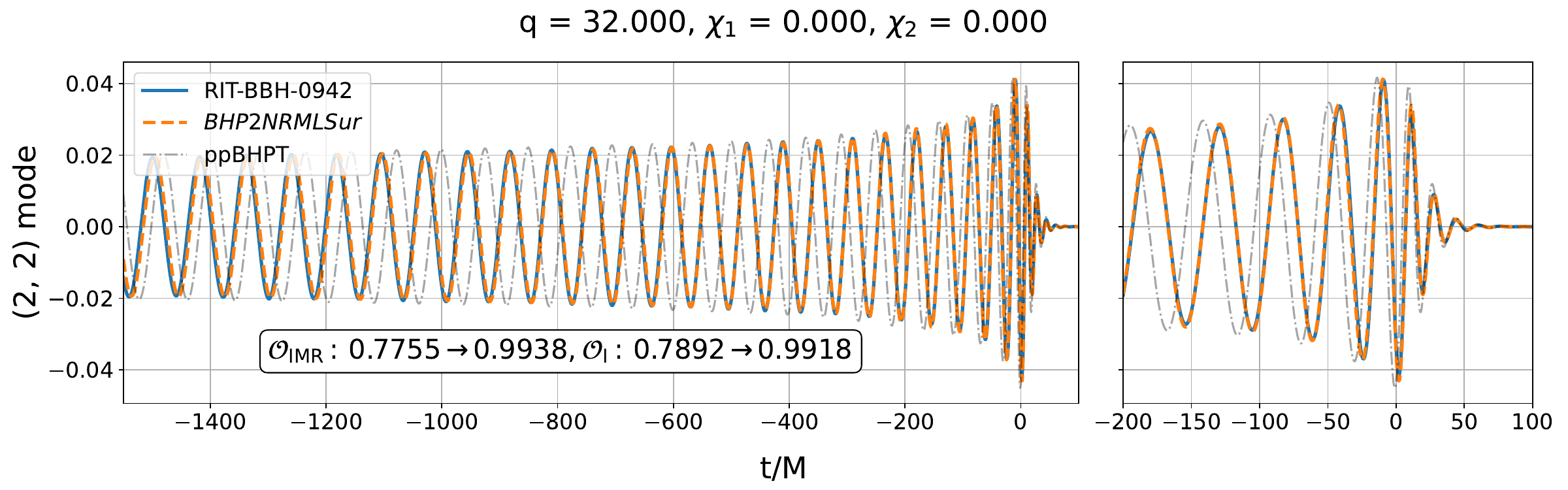}
\captionsetup{justification=raggedright, singlelinecheck=false}
\caption{Waveforms generate by \textit{BHP2NRMLSur} 
(orange line) and NR simulations (blue line) from the RIT group (simulation ID RIT: BBH:0942 for $q = 15$ and RIT: BBH:0792 for $q = 32$). The uncalibrated ppBHPT waveforms are shown in light black dashed lines. The matches of the waveforms before and after mapping and NR simulations are marked in the text box. } \label{RIT}
\end{figure}

We compare the \textit{BHP2NRMLSur}  waveforms and RIT NR simulations \cite{RIT2020} for the high mass-ratio cases $q=\{15,\,32\}$. Figure~\ref{RIT} shows the comparison against RIT NR data. The matches are $0.9973$ and $0.9938$ corresponding to $q=15$ and $q=32$. Since the mass ratio of the training data is sampled in the range $q\in[3,\,8]$, the results show that the nonspinning \textit{BHP2NRMLSur} model can be generalized to out-of-distribution mass ratios and remains valid for larger mass-ratio cases.


In this work, the ppBHPT waveforms we use as part of the input training data are for nonspinning systems. In the future, we will consider using ppBHPT waveforms with more parameters, such as spin and eccentricity $e$ to generate a more general mapping model as
\begin{equation}
{\rm CfC}^{lm}_\alpha[\alpha(q,\,\chi,\,e,\,...)]=\left(1+\sum_{n=1}^{n_{\rm max}}\frac{{\rm CfC}^{lmn}_\alpha}{q^{n}}\right)\alpha(q,\,\chi,\,e,\,...).\label{fu}
\end{equation}

\subsection{Aligned-spinning model}
The aligned-spinning \textit{BHP2NRMLSur} model maps nonspinning ppBHPT into aligned-spinning waveforms via
\begin{equation}
\alpha(q,\chi_1,\chi_2)={\rm CfC}^{lm}_{\alpha}[\alpha(q),\,\chi_1 ,\,\chi_2],\label{chi12}
\end{equation}
where the CfC network has $64$ neurons ($N_{\rm tot}=64$). Note that unlike Eq.~(\ref{fu}), $\alpha(.)$ in Eq.~(\ref{chi12}) has only one parameter, mass-ratio $q$. This means that the inputs to the \textit{BHP2NRMLSur} are the nonspinning waveforms and the two spin parameters $\chi_1$ and $\chi_2$, and the outputs of the model obtained are the aligned-spinning waveforms.

Here, we generate the NRHybSur3dq8$\_$CCE-based model and the SEOBNRv5HM-based model depending on whether the target data in training are from NRHybSur3dq8$\_$CCE or SEOBNRv5HM, respectively. For the NRHybSur3dq8$\_$CCE-based model, the input ppBHPT data $\alpha(q)$ is the amplitude and phase of 30 waveforms with uniformly sampled mass-ratio $q\in[3,8]$. The input parameters are $30\times30$ uniformly distributed spins of $\chi_1\in[-0.8,0.8]$ and $\chi_2\in[-0.8,0.8]$. The dimension of the target data for the corresponding parameters is $30\times30\times30$. Since we have chosen the SEOBNRv5HM target data with a larger parameters (mainly the mass-ratio) interval, we increase the dimension of the training data moderately. In the training of the SEOBNRv5HM-based mode, the dimension of target data is $50\times50\times50$, with $50$ logarithmically sampled mass-ratio  $q\in[1,200]$, and $50\times50$ uniformly sampled spins of $\chi_1\in[-0.9,0.9]$ and $\chi_2\in[-0.9,0.9]$. In addition, we test the case when the dimension is $30\times30\times30$ and find that the model cannot provide good accuracy, $50$ mass-ratio sampling is necessary.

\begin{figure}[htb!]
\centering
\includegraphics[width=0.49\textwidth]{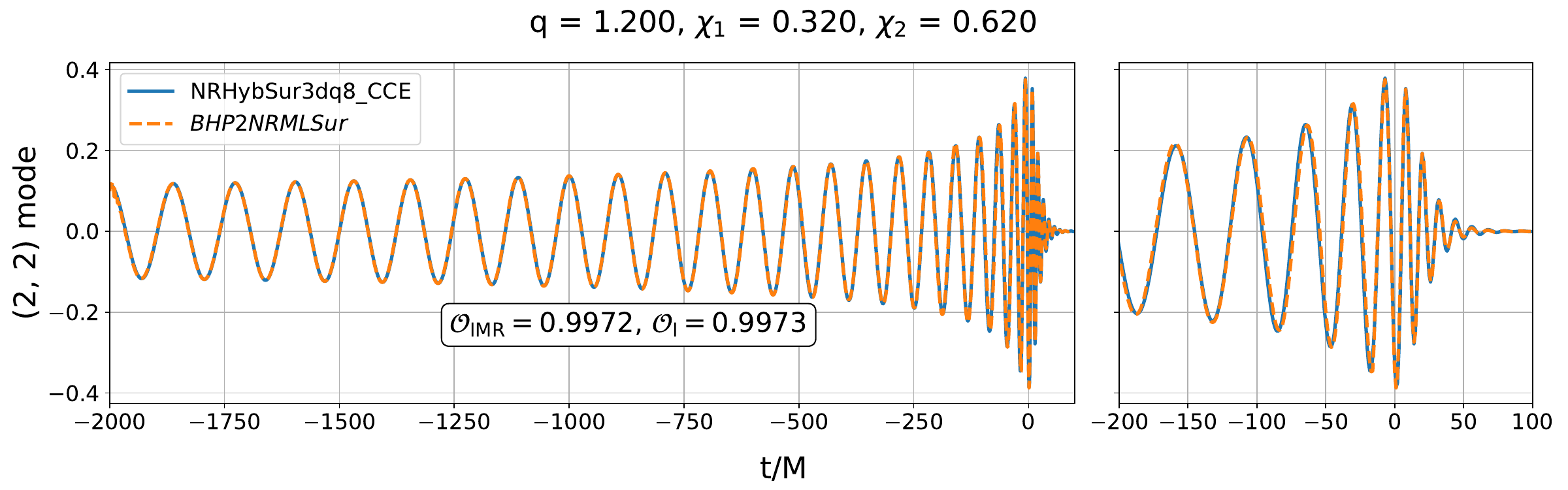}
\includegraphics[width=0.49\textwidth]{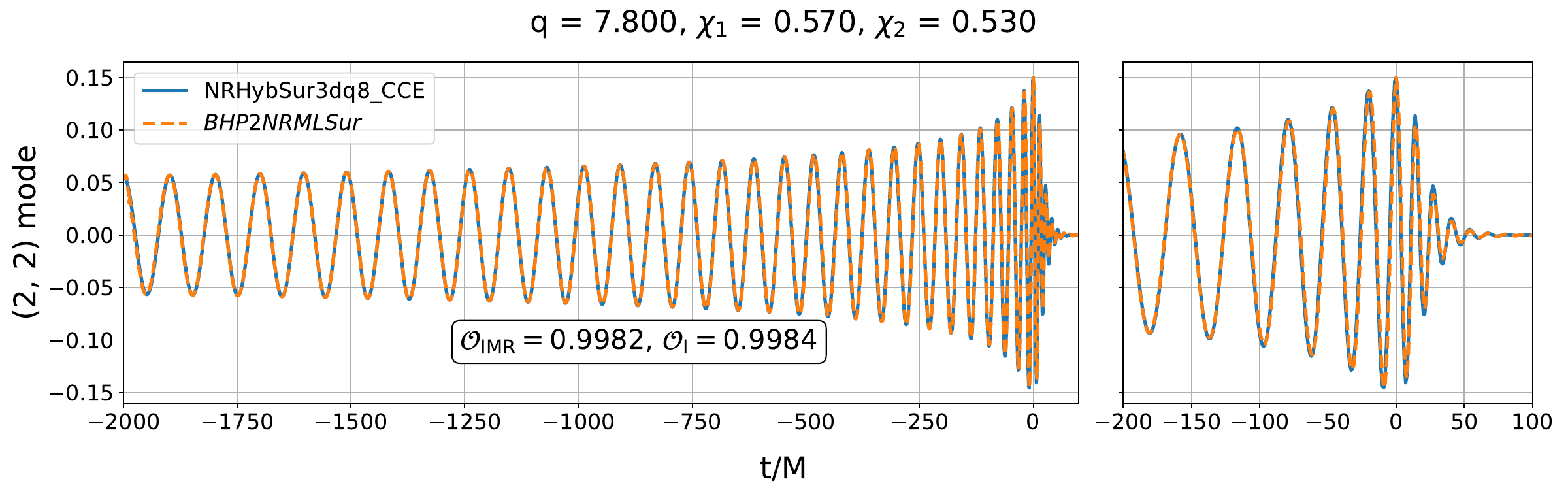}
\captionsetup{justification=raggedright, singlelinecheck=false}
\caption{Waveforms generate by \textit{BHP2NRMLSur} $\rm NRHybSur3dq8\_CCE$-based model and $\rm NRHybSur3dq8\_CCE$, where the \textit{BHP2NRMLSur} waveforms are represented by orange dashed lines and the corresponding $\rm NRHybSur3dq8\_CCE$ waveforms are represented by solid blue lines.} \label{cce}
\end{figure}

\begin{figure}[htb!]
\centering
\includegraphics[width=0.49\textwidth]{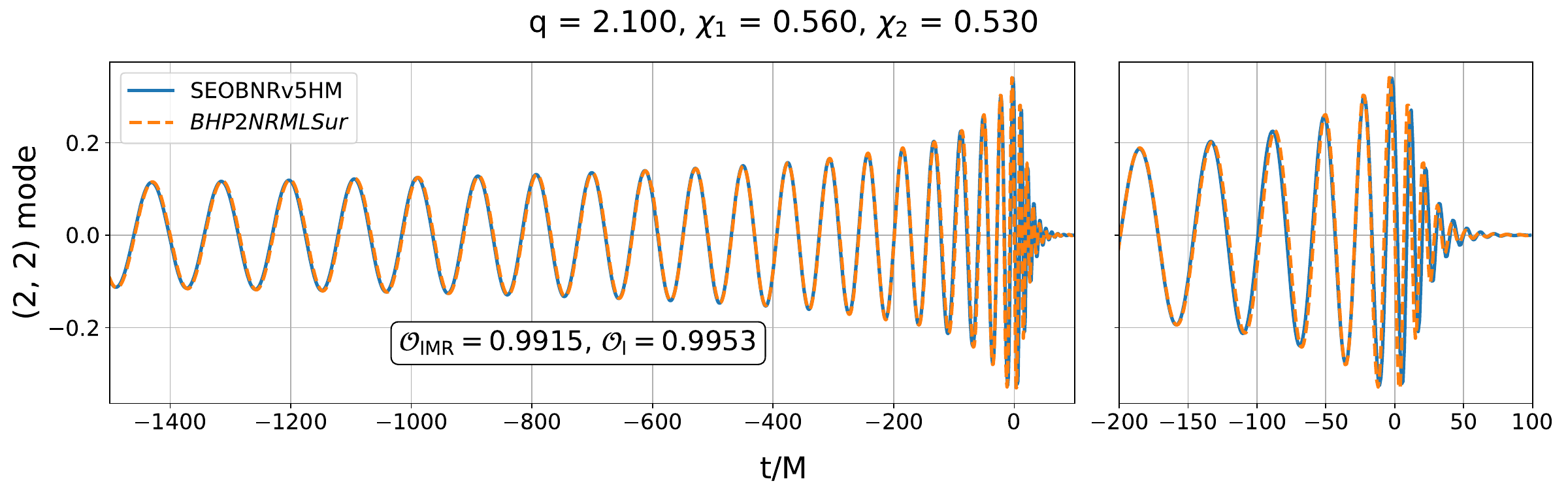}
\includegraphics[width=0.49\textwidth]{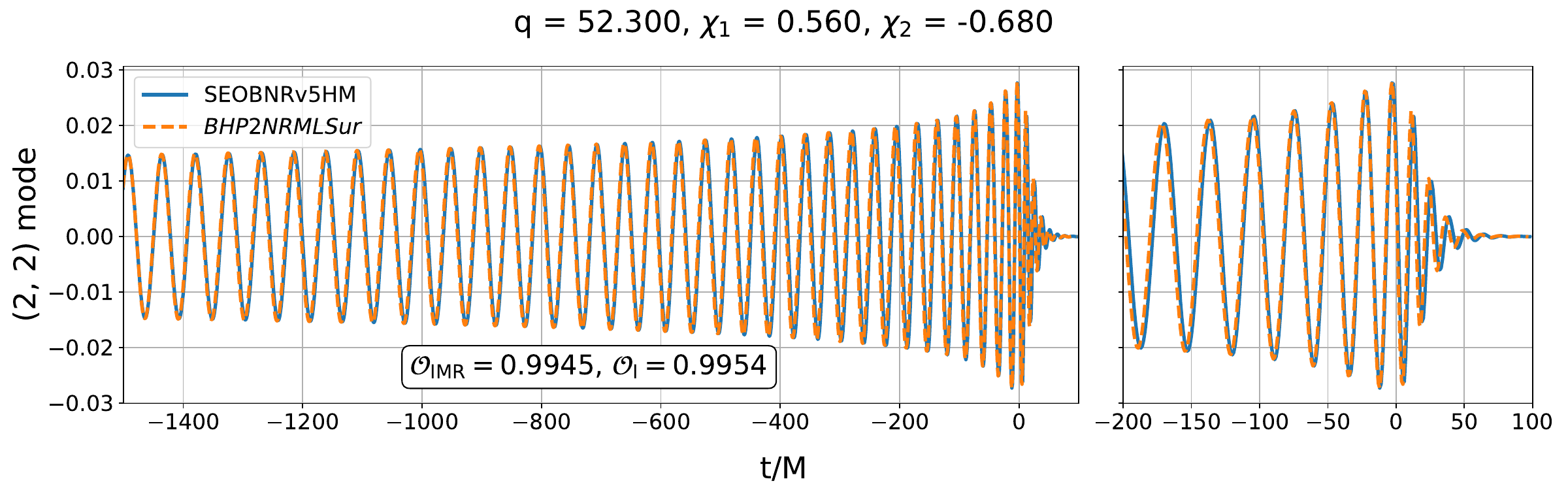}
\captionsetup{justification=raggedright, singlelinecheck=false}
\caption{Waveforms generate by \textit{BHP2NRMLSur} SEOBNRv5HM-based model and SEOBNRv5HM, where the \textit{BHP2NRMLSur} waveforms are represented by orange dashed lines and the corresponding SEOBNRv5HM waveforms are represented by blue solid lines.} \label{seob}
\end{figure}

\begin{figure}[htb!]
\centering
\hspace*{-0.22cm}
\includegraphics[width=0.49\textwidth]{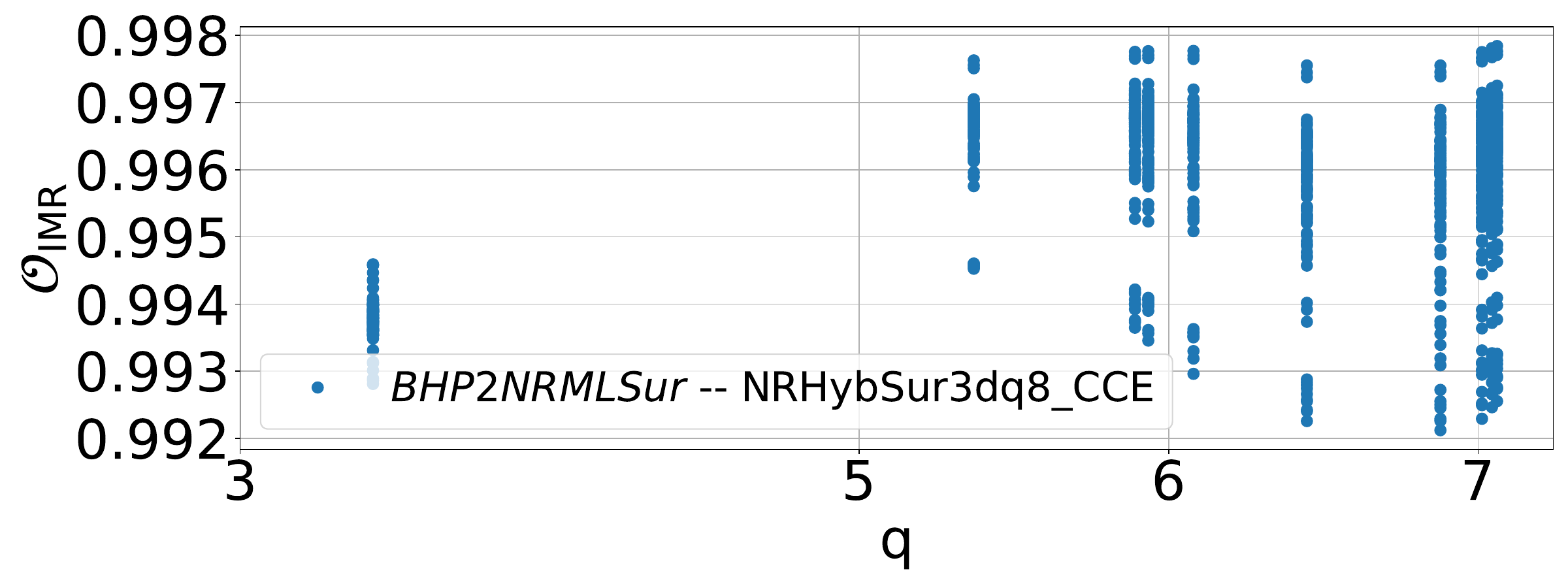}
\includegraphics[width=0.49\textwidth]{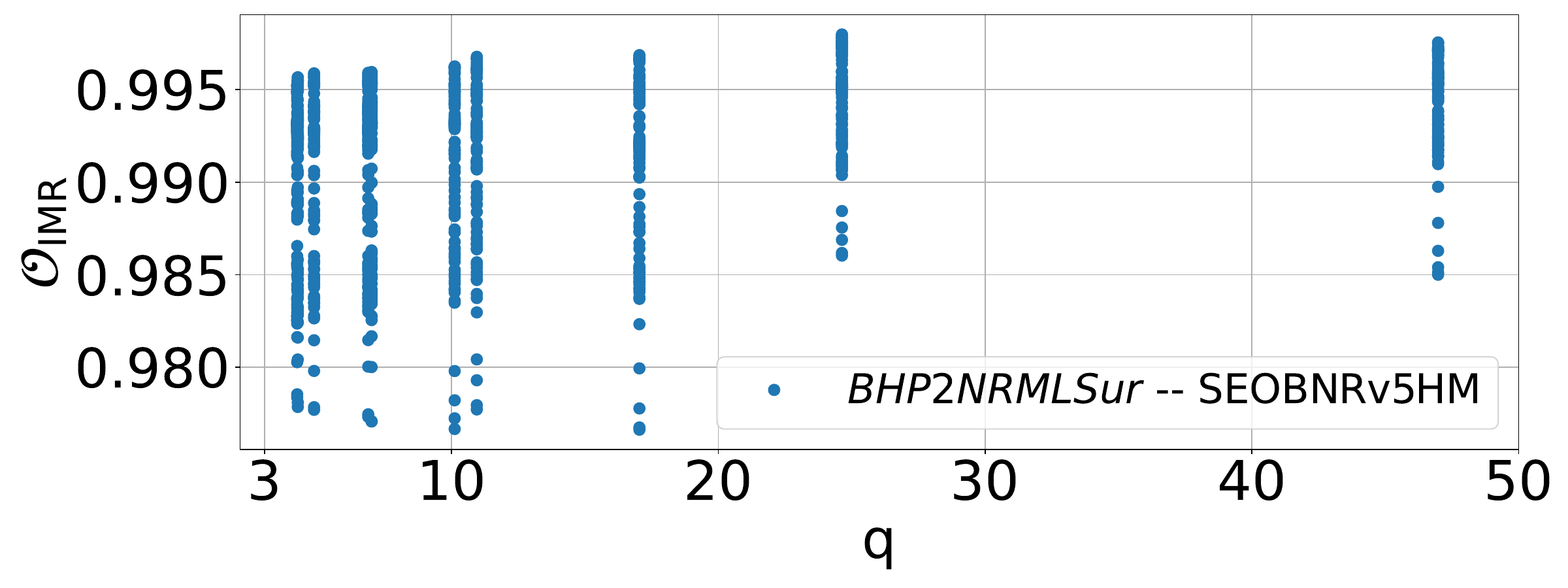}
\captionsetup{justification=raggedright, singlelinecheck=false}
\caption{The matches $\mathcal{O}_{\rm IMR}$ of \textit{BHP2NRMLSur} waveforms and $\rm NRHybSur3dq8\_CCE$(top)/ SEOBNRv5HM (bottom). Both panels have $1000$ waveform matches, and each column of $100$ points corresponds to matches of the same mass-ratio $q$ and different spins.} \label{seob-bhp-cce}
\end{figure}

Similar to the nonspinning case, we generate waveforms using our \textit{BHP2NRMLSur}  models and calculate the matches to quantify the model accuracy using Eq.~(\ref{meq}). Figs.~\ref{cce} and \ref{seob} show the examples of waveforms generated by \textit{BHP2NRMLSur} using the NRHybSur3dq8$\_$CCE-based and SEOBNRv5HM-based models, respectively. The match results are shown in Figure~\ref{seob-bhp-cce}, where the results for each model are computed from $1000$ waveforms. For the NRHybSur3dq8$\_$CCE-based model, a total number of $10\times10\times10$ data points are randomly drawn from a uniform distribution, with $q \in [3,8]$, $\chi_1 \in [0, 0.7]$, and $\chi_2 \in [0, 0.7]$. As for the SEOBNRv5HM-based model, $10$ mass-ratio values are randomly drawn from a uniform distribution over the range $q\in[3, 50]$, while the $10\times10$ spin values are randomly drawn from a uniform distribution with $\chi_1\in[0, 0.7]$ and $\chi_2\in[- 0.7,0.7]$. The matches of the waveforms generated by the NRHybSur3dq8$\_$CCE-based and SEOBNRv5HM-based models are above $0.996$ and $0.97$, respectively.

\begin{figure}[htb!]
\centering
\includegraphics[width=0.47\textwidth]{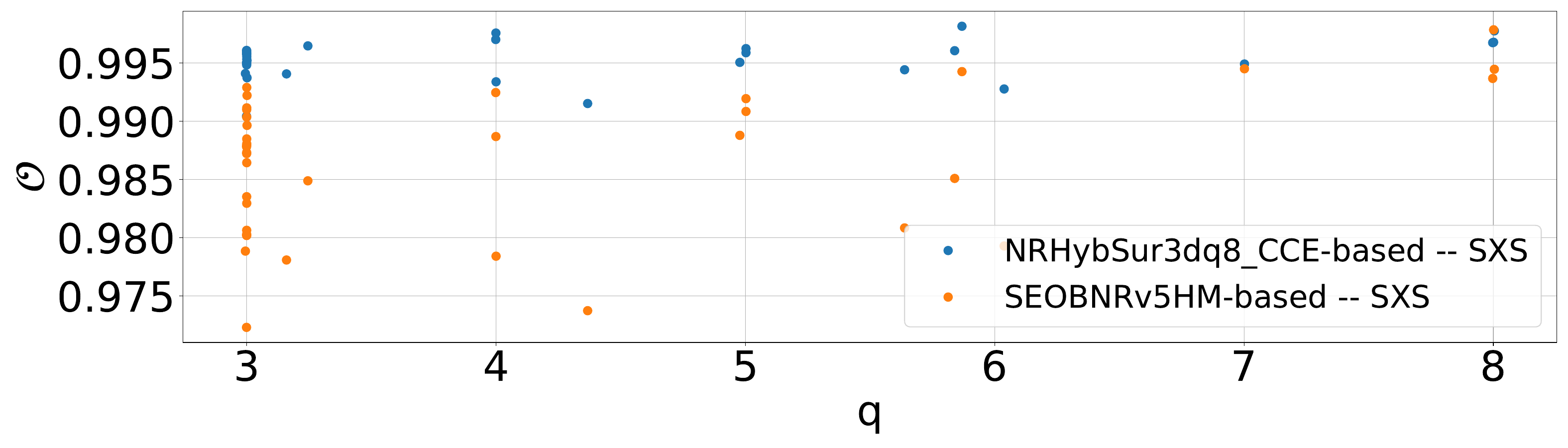}
\captionsetup{justification=raggedright, singlelinecheck=false}
\caption{The matches $\mathcal{O}_{\rm IMR}$ of \textit{BHP2NRMLSur} waveforms and SXS simulation. The blue points indicate the matches of the NRHybSur3dq8$\_$CCE-based model case, and the orange points correspond to the SEOBNRv5HM-based model case.} \label{SXS-l}
\end{figure}

We also compare the waveforms generated by the \textit{BHP2NRMLSur} model to NR simulation. Here, we select $30$ waveforms with the aligned-spinning $\chi_1\in [0, 0.85]$ and $\chi_2 \in[0,0.85]$ and eccentricity $e\leq 0.01$ from the SXS GW database. Figure~\ref{SXS-l} shows the matches between the \textit{BHP2NRMLSur} and SXS waveforms. The matches are above $0.99$ and 0.97 for the waveforms generated by the NRHybSur3dq8$\_$CCE-based model and SEOBNRv5HM-based model, respectively.

\subsection{Generation efficiency}
Since the input ppBHPT waveforms have only one parameter $q$, the \textit{BHP2NRMLSur} model uses the nonspinning waveforms as the basis to obtain aligned-spinning waveforms with $\chi_1$ and $\chi_2$. This model reduces the dimension of the input to achieve improved computational efficiency, tens of times faster than the surrogate models $\rm NRHybSur3dq8\_CCE$ and SEOBNRv5HM by using the GPU RTX A2000 12GB.


\begin{figure}[htb!]
\centering
\includegraphics[width=0.45\textwidth]{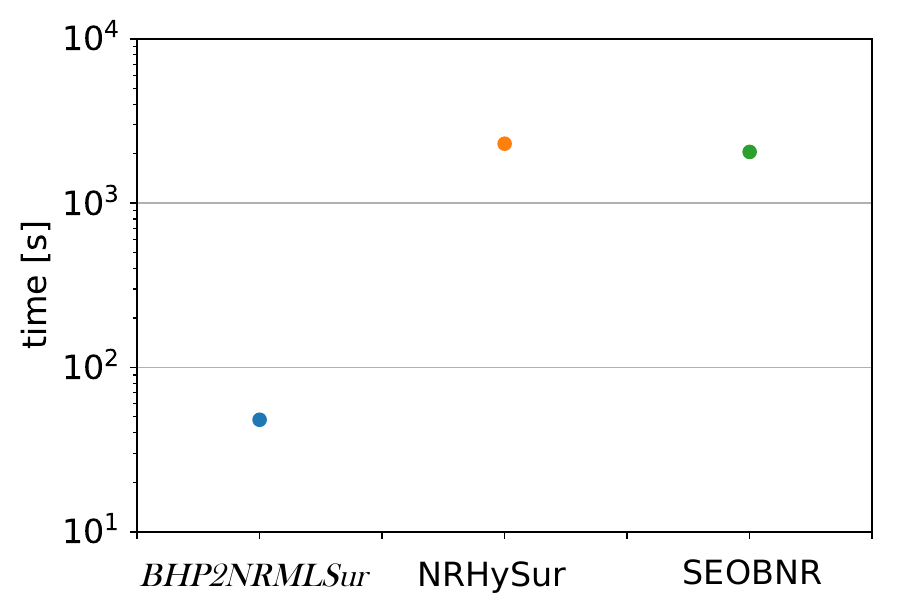}
\captionsetup{justification=raggedright, singlelinecheck=false}
\caption{Computation time for generating $100000$ waveforms by different waveform models, where the blue, orange, and green points represent \textit{BHP2NRMLSur}, $\rm NRHybSur3dq8\_CCE$ and SEOBNRv5HM respectively.} \label{speed}
\end{figure}

In Figure~\ref{speed}, we show the computing time for generating $100$-thousand waveforms by $\rm NRHybSur3dq8\_CCE$, SEOBNRv5HM, and our \textit{BHP2NRMLSur} model. The results demonstrate the advantage of the \textit{BHP2NRMLSur} model. The computing time of our model is about $50$ seconds, $\sim 40$ times faster than using $\rm NRHybSur3dq8\_CCE$ or SEOBNRv5HM, which takes $\sim 2\times 10^3$ seconds. 

\subsection{Scaling parameters extraction and comparison}
Similar with Refs.~\cite{rifat2020surrogate,islam2022surrogate, PhysRevD.110.124069}, we can also extract the parameters in Eq.~(\ref{alpha_beta}) from our machine-learning model and compare them with the phenomenological formulae used in the above works. Here, we take the nonspinning case as an example. The definition of the scaling form is as follows
\begin{equation}
{h}^{\alpha_l'\beta'}_{lm}(t,\,q) = \alpha'_l{A}(t)e^{-i\phi(t,\,q)/\beta'},
\end{equation}
where ${h}^{\alpha_l'\beta'}_{lm}(t,\,q)$ is the waveform after the calibration with scaling parameters. Although $\beta'$ scales the phase, different from $\beta$ in Reference~\cite{islam2022surrogate} which scales the time, they are essentially equivalent since one operation is multiplication and the other is division.

Now, we fit the coefficients $(A,\, B,\, C,\, D)$ of $\alpha_l'$ and $\beta'$ by using the least-squares method. The values of these coefficients are presented in Table~\ref{alpha_T} and~\ref{beta_T}. The coefficients obtained here are similar in terms of the trend with the ones in ~\cite{islam2022surrogate},  with slight differences. Moreover, the relative differences between ($\alpha_l$, $\beta$) and ($\alpha_l'$, $\beta'$) just from 1\% to 10\% for varied mass-ratio. The scaling formulae based on ($\alpha'_l$, $\beta'$) can effectively rescale the machine-learning waveforms, especially for the $(2,\,2)$ mode, however for higher modes, the accuracy may be not enough.

\begin{figure}
\centering
\includegraphics[width=0.49\textwidth]{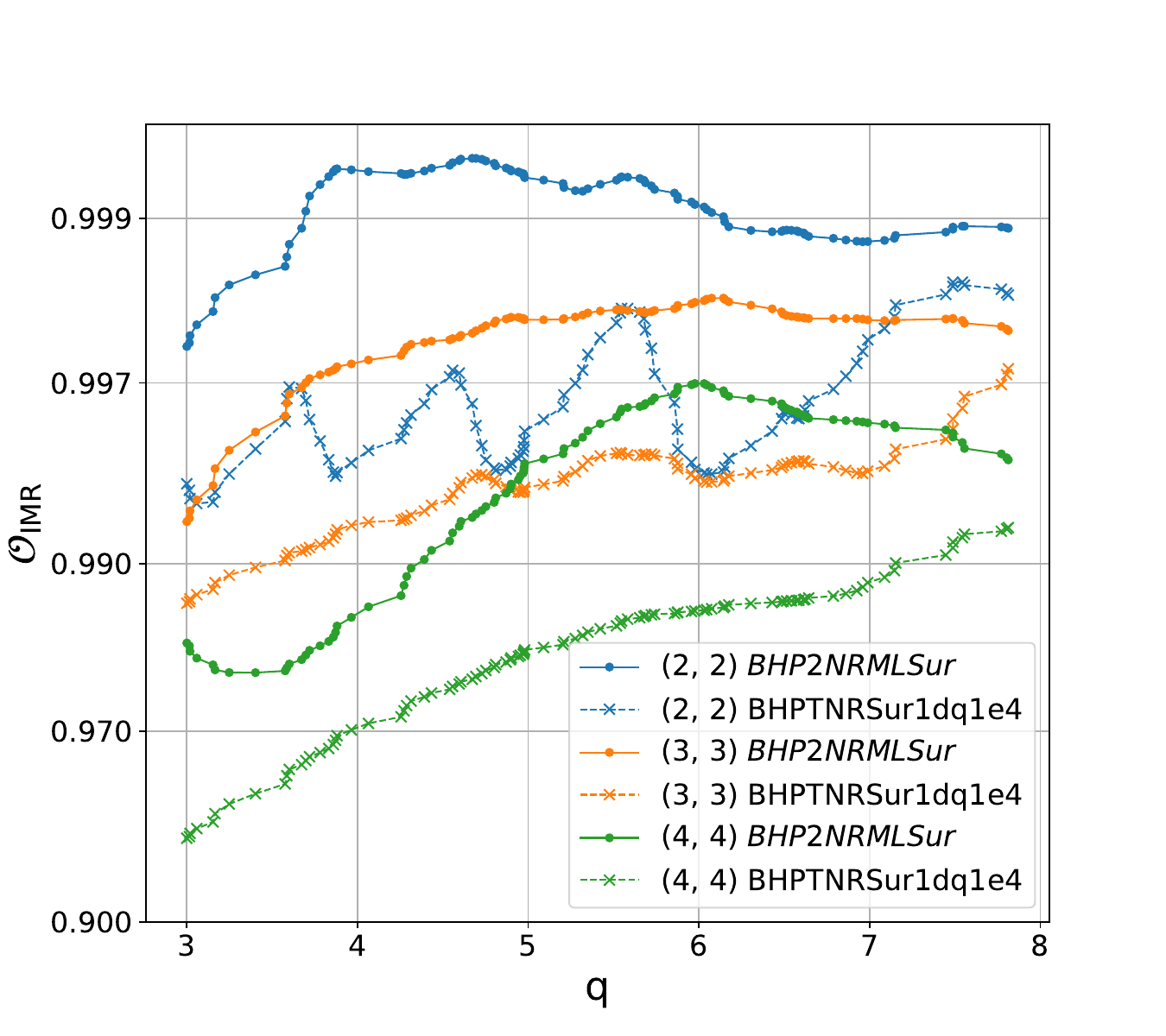}
\captionsetup{justification=raggedright, singlelinecheck=false}
\caption{Matches ($\mathcal{O}_{\rm IMR}$) by comparing NRHybSur3dq8$\_$CCE waveforms to \textit{BHP2NRMLSur} and BHPTNRSur1dq1e4 waveforms.} \label{calmatch}
\end{figure}

Based on the above analysis, it is better to use the machine-learning waveform directly. We now compare the accuracy between our \textit{BHP2NRMLSur} and the BHPTNRSur1dq1e4 model from Reference~\cite{islam2022surrogate}. For the $(2,\,2)$, $(3,\,3)$, and $(4,\,4)$ waveform modes, we compute the matches $\mathcal{O}_{\rm IMR}$ for both \textit{BHP2NRMLSur} and BHPTNRSur1dq1e4 against NRHybSur3dq8$\_$CCE. As shown in Figure~\ref{calmatch}, \textit{BHP2NRMLSur} demonstrates better accuracy across all three modes. Furthermore, \textit{BHP2NRMLSur} exhibits strong parametric scalability, enhancing its potential for further development and application. 

\begin{table}[htb!]
\captionsetup{justification=raggedright, singlelinecheck=false}
\caption{\label{alpha_T}Fitting coefficients for $\alpha'_l$ parameters as defined in Eq.~(\ref{alpha_l}).}
\centering
\setlength\tabcolsep{1.3pt}
\begin{tabular}{c|c|c|c|c}
\hline\hline
 $l$&$A^l_\alpha$&$B^l_\alpha$&$C^l_\alpha$&$D^l_\alpha$\\
 \hline
 2&$-1.488\pm0.006$&$2.341\pm0.075$&$-4.911\pm0.327$&$6.302\pm0.459$
 \\
 \hline
 3&$-3.243\pm0.006$&$8.138\pm0.078$&$-27.651\pm0.338$&$44.888\pm0.476$
 \\
 \hline
 4&$-3.724\pm0.013$&$5.615\pm0.176$&$-1.306\pm0.768$&$-8.789\pm1.080$
 \\
\hline\hline
\end{tabular}
\end{table}

\begin{table}[H]
\caption{\label{beta_T}Fitting coefficients for $\beta'$ parameters as defined in Eq.~(\ref{beta_l}).}
\centering
\begin{tabular}{c|c|c|c}
\hline\hline
 $A_\beta$&$B_\beta$&$C_\beta$&$D_\beta$\\
 \hline
 $-0.944 \pm0.0003$&$1.618 \pm 0.004$&$-2.277 \pm 0.018 $&$1.457 \pm 0.027$
 \\
\hline\hline
\end{tabular}
\end{table}



\section{Discussion}
\label{sec:discussion}
In this work, we present a new machine-learning waveform mapping strategy that can calibrate the ppBHPT waveforms to NR surrogate waveforms. We obtain the \textit{BHP2NRMLSur} based on the CfC networks in Eq.~(\ref{CfCEq}) with the multiple-neuron structure NCP, and training with the ppBHPT and NR surrogate (SEOBNRv5HM and NRHybSur3dq8$\_$CCE) waveform data. This method converts nonspinning ppBHPT waveforms to NR surrogate waveforms with the advantages of high accuracy, high speed, and parameter scalability. The main contribution of our work is the first use of a machine-learning approach to handle waveform mapping from the ppBHPT to NR surrogate with theoretically arbitrary parameter scalability.  

Based on the proposed method, we train to obtain \textit{BHP2NRMLSur} with two mapping models: nonspinning and aligned-spinning models. The nonspinning model can generalize the waveforms to out-of-distribution mass ratios. The waveforms generated by the trained \textit{BHP2NRMLSur} model are highly accurate compared to the NR surrogate waveforms. In particular, for the first three main modes $(2,~2)$, $(2,~1)$ and $(3,~3)$, the matches $\mathcal{O}_{\rm IMR}$ of waveforms generated by \textit{BHP2NRMLSur} and NRHybSur3dq8$\_$CCE are above $0.99$. For mass ratio $q>5$, matches $\mathcal{O}_{\rm IMR}>0.99$ are also achieved in the $(3,~2)$ and $(4,~4)$ modes. For the aligned-spinning model, by mapping nonspinning ppBHPT waveforms to NR surrogate waveforms with aligned-spinning $\chi_1$ and $\chi_2$, we take advantage of parameter space reduction and generate waveforms tens of times faster than current surrogate models, SEOBNRv5HM and NRHybSur3dq8$\_$CCE. In future work, we will train the aligned-spinning ppBHPT waveform data using the model of Eq.~(\ref{fu}), which will make the out-of-distribution mass ratios of the aligned-spinning model possible.

Our machine-learning framework for calibrating ppBHPT waveforms to NR surrogate waveforms is generalized and expected to apply to a variety of scenarios not demonstrated in this work, including a variety of other time-series calibrations such as other waveform calibration and the calibration of orbital parameters evolution, etc. Compared with existing polynomial-fitting modulation methods, the \textit{BHP2NRMLSur} model exhibits high-parameter precision and strong scalability. Furthermore, with more NR simulation data in the future, the model can be extended to more general cases, such as eccentric and precessing systems.


\section{Acknowledgements}
This work is supported by the National Key R\&D Program of China (Grant No. 2021YFC2203002), and the National Natural Science Foundation of China (Grants No. 12173071, 12473075). Xing-Yu Zhong was supported by the UCAS to participate in a Joint PhD Training Program. This work made use of the High-Performance Computing Resource in the Core Facility for Advanced Research Computing at Shanghai Astronomical Observatory. L.S. acknowledges the Australian Research Council Centre of Excellence for Gravitational Wave Discovery (OzGrav), Projects No. CE170100004 and No. CE230100016. L.S. is also supported by the Australian Research Council Discovery Early Career Researcher Award, Project Number DE240100206.

\bibliography{groupref}
\bibliographystyle{apsrev4-1}
\end{document}